\documentclass[aps,prb,twocolumn,superscriptaddress]{revtex4-2}

\usepackage{graphicx}
\usepackage{amsmath, bm}
\usepackage{amssymb}
\usepackage{lineno}
\usepackage{ulem}
\usepackage{color}

\newcommand{\val}[2]{$#1 \times 10^{#2}$}

\usepackage{siunitx}

\begin{document}
\title{
Impact of Spin Wave Dispersion on Surface Acoustic Wave Velocity
}

\author{Pauline Rovillain}
\email{pauline.rovillain@insp.upmc.fr}
\affiliation{Sorbonne Universit\'e, CNRS, Institut des NanoSciences de Paris, INSP, UMR7588, F-75005 Paris, France}
\author{Jean-Yves Duquesne}
\affiliation{Sorbonne Universit\'e, CNRS, Institut des NanoSciences de Paris, INSP, UMR7588, F-75005 Paris, France}
\author{Louis Christienne}
\affiliation{Sorbonne Universit\'e, CNRS, Institut des NanoSciences de Paris, INSP, UMR7588, F-75005 Paris, France}
\author{Mahmoud Eddrief}
\affiliation{Sorbonne Universit\'e, CNRS, Institut des NanoSciences de Paris, INSP, UMR7588, F-75005 Paris, France}
\author{Maria Gloria Pini}
\affiliation{ Istituto dei Sistemi Complessi del CNR (CNR-ISC), Sede Secondaria di Firenze, I-50019 Sesto Fiorentino, Italy}
\author{Angelo Rettori}
\affiliation{Dipartimento di Fisica ed Astronomia, Universit\'a degli Studi di Firenze, I-50019 Sesto Fiorentino, Italy}
\affiliation{INFN, Sezione di Firenze, 50019 Sesto Fiorentino, Italy}
\author{Silvia Tacchi}
\affiliation{Istituto Officina dei Materiali del CNR (CNR-IOM), Sede Secondaria di Perugia, c/o Dipartimento di Fisica e Geologia, Universit\'a di Perugia, I-06123 Perugia, Italy}
\author{Massimiliano Marangolo}
\affiliation{Sorbonne Universit\'e, CNRS, Institut des NanoSciences de Paris, INSP, UMR7588, F-75005 Paris, France}

\date{\today}

\begin{abstract}
The dependence of the velocity of surface acoustic waves (SAWs) as a function of an external applied magnetic field is investigated in a  Fe thin film epitaxially grown on a piezoelectric GaAs substrate. The SAW velocity is observed to strongly depend on both the amplitude and direction of the magnetic field.  To interpret the experimental data  a phenomenological approach  to the relative change in SAW velocity is implemented. We find that the experimental velocity variation can be well reproduced provided that the spin wave dispersion is taken into account. The validity of this  phenomenological model is attested by the comparison with a fully magnetoelastic one. Non-reciprocity of SAW velocity is also addressed both experimentally and theoretically.
\end{abstract}

\maketitle

\section{Introduction}

In the last years, the coupling between surface acoustic waves (SAWs) and spin waves (SWs) \cite{Lum1958} in magnetic films has been the subject of increasing interest due to the possible application in the framework of magnonics \cite{Chumak2015,Barman2021}. This latter research field aims to process information at low power consumption, small footprint and high operation speed in a wide frequency range from GHz to THz.
In this context, the SAWs have been proposed to dynamically control SW propagation and even to generate SWs, in order to implement reconfigurable and energy efficient magnonic devices \cite{Barman2021}. 
Indeed, SAW technology is mature and widely used in today's sensors, filters and microwave circuitry, notwithstanding the lack of tunability of SAW transducers. For this reason a multitude of them are currently integrated in modern devices (e.g. mobile phones). In this context, it has been shown that tunability can be increased by using electric \cite{Li2018} or magnetic \cite{Zhou2014} fields. 
\\
Recently, the so-called SAW induced Ferromagnetic Resonance (SAW-FMR) has been observed by Weiler \textit{et al.} \cite{Weiler2011}, Thevenard \textit{et al.} \cite{Kuszewski2018}, and Duquesne \textit{et al.} \cite{Duquesne2019a} in Ni, GaMnAs and Fe thin films, respectively, by exciting SAWs in the GHz and sub-GHz regime in piezoelectric media. Moreover, it is worthwhile to notice that SAW-FMR permits to reverse magnetization \cite{Thevenard2016} and even to induce spin-pumping in  CoPt bi-layers \cite{Weiler2012, Rovillain2020}.

The SAW-FMR interaction is often described \cite{Dreher2012} by taking into account only the uniform FMR mode, i.e. $k_{\rm SW}~=~0$. 
However this approximation is rather crude, and it misses out the wealth of modes that can be excited in a ferromagnetic (FM) material. 
Moreover, when compared to other waves (photons, phonons, etc...), SWs present two important features: their frequency can be easily varied by applying an external magnetic field $\mathbf{B}_{\rm ext}$, and their dispersion curves strongly depend on the direction
of $\mathbf{B}_{\rm ext}$ with respect to the SW wavevector $\mathbf{k}$. Because of long-range magnetic dipole-dipole interactions, SWs at low wavevectors present negative (positive) group velocity when the magnetization is parallel (perpendicular) to the $\mathbf{k}$-vector, corresponding to the so-called Backward (Damon-Eshbach) configuration \cite{review}. Therefore the in-plane anisotropy of SWs dispersion has to be taken into account for the design of magnonic devices where SAWs are exploited to control or generate SWs.
Interestingly, some recent articles put into evidence the important role played by a nonzero wavevector \cite{Gowtham2015, Hernandez-Minguez2020, Babu2021, Xu2020, Verba2018, Kuss2021}. \\

In particular, Babu \textit{et al.} studied the SAW-SW coupling by Brillouin Light Scattering (BLS) in a wide $\mathbf{k}$-values range \cite{Babu2021}. They showed that in order to obtain an appreciable coupling between SWs and SAWs, the SAW profile within the magnetostrictive layer is also of importance, \cite{Babu2021} in addition to the anisotropy of the magnetoelastic interaction with respect to the orientation of the magnetic field \cite{Dreher2012}.

In our previous work \cite{Duquesne2019a} we have shown that SAW-FMR can be obtained in an epitaxial Fe thin film by monitoring attenuation and velocity changes of the SAWs. Here, we investigate the dependence of the SAW velocity variation $\Delta V/V$ as a function of the in-plane direction of the external applied magnetic field $\mathbf{B}_{\rm ext}$. We find that $\Delta V/V$ strongly depends on the orientation between $\mathbf{B}_{\rm ext}$ and the SAW wave-vector ($\mathbf{k}_{\rm SAW}$). In particular, we show that in order to well reproduce and interpret the experimental results, the SWs dispersion has to be included in the theoretical model. Our approach is an extension of the model given by Dreher \textit{et al.} \cite{Dreher2012} to describe the backaction of the ferromagnetic resonance on the acoustic wave. In literature, this phenomenological model has been adopted to quantify the part of the SAW power used to drive the magnetization dynamics by considering SAW propagation and FMR in the ferromagnetic thin film, i.e. by neglecting SAW propagation in the substrate \cite{Duquesne2019a,Dreher2012,Hernandez-Minguez2020}.
Here, we show that the backaction model is also able to catch the physics of SAW velocity changes in a ferromagnetic thin film. Moreover, we corroborate this approximate approach by a comparison with a more sophisticated calculation, i.e. a fully magnetoelastic approach, which considers SAW propagation in the heterostructure as a whole, i.e. composed of the thin magnetic layer on a substrate.

A full comprehension of SAW-SW interaction opens up dazzling Janus-faced applications: on one side, a single IDT can provide the energy needed to activate magnetization dynamics 
in a Joule heat free manner. On the other side, the too limited phase and/or resonance frequency tunability in SAW filters can be augmented by an external magnetic field, as shown by Zhou \textit{et al.} \cite{Zhou2014}.

\section{Samples characteristics and experimental setup}

\begin{figure}
    \centering
    \includegraphics[width=\linewidth]{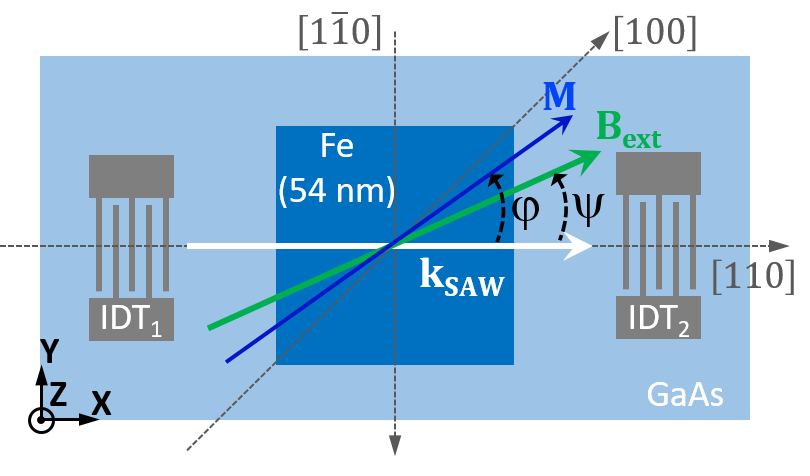}
    \caption{Sketch of the sample: $4 \times 4$ mm$^2$ Fe mesa (54~nm thick) on GaAs(001). The IDT has a split-44 design\cite{Schulein2015}. The wave vector $\mathbf{k}_{\rm SAW}$ is parallel to $[110]$, $\mathbf{B}_{\rm ext}$ is the in-plane applied magnetic field, and \textbf{M} is the Fe magnetization. }
    \label{fig:dispo}
\end{figure}

A single-crystal Fe thin film with a thickness of 54 nm \footnote{ The thickness of 67~nm reported in \cite{Duquesne2019a} 
was the nominal one. After examination and comparison with BLS data (see in particular Fig.~\ref{fig:B0} in Appendix B), the actual thickness was found to be 54~nm.} was deposited on a ZnSe(001) epilayer ($\approx$ 20~nm) grown on GaAs(001) substrate. An Au capping layer ($\approx$ 8~nm) was deposited to protect the Fe thin film. The layer thickness was determined by TEM and X-ray reflectivity. The heterostructure exhibits the expected epitaxial growth relationship, which is [100]Fe//[100]ZnSe//[100]GaAs and (001)Fe//(001)ZnSe//(001)GaAs. In agreement with previous work \cite{Marangolo}, vibrating sample magnetometer measurements show that the sample is characterized by in-plane biaxial magnetic anisotropy, typical of the bulk Fe, with the magnetic easy and hard axis along the $[100]$ and the $[110]$ directions, respectively. Details of the growth, characterization, and magnetic properties of Fe thin films are given in Ref.~\onlinecite{Duquesne2019a}. 
Then a $4 \times 4$ mm$^2$ mesa is defined by ion etching in order to perform acoustic measurements.
SAWs were excited along the $[110]$ direction ($\mathbf{k}_{\rm SAW}$~//~$[110]$) at four harmonic frequencies $\nu_{SAW} \approx q \nu$, with $\nu$~=~119~MHz and $q$ = 1, 3, 5, 7, by a ``split-44" interdigitated transducer (IDT) \cite{Schulein2015}. 
An identical IDT acts as a detector. The $\mathbf{k}_{\rm SAW}$ direction can be reversed by connecting the input signal to one IDT or to the other.  Pulsed excitation was used with 500~ns duration and the acoustic signal is acquired with a digital oscilloscope to do direct time domain measurements \cite{Thevenard2014}. A sketch of the device is given in Fig.~\ref{fig:dispo}. Prior to each measurement, the sample was saturated by a large in-plane external reference magnetic field $\mathbf{B}_{\rm ext}$, having magnitude $B_{\rm ref}$ = 0.4~T and forming an angle $\psi$ with the [110] direction. The change of the relative SAW velocity $\Delta V/V = [V(B_{\rm ext}) - V(B_{\rm ref})]/V(B_{\rm ref})$ was obtained by measuring the phase of the acoustic signals \footnote{All the results are shown with a reference field at saturation to be sure of the magnetic state. At high fields, the velocity values differ, depending on the frequency and the magnetization vector direction. Consequently, at zero field the phase shift for all measurements do not overlap.}. 

\begin{figure}
    \centering
    \includegraphics[width=\linewidth]{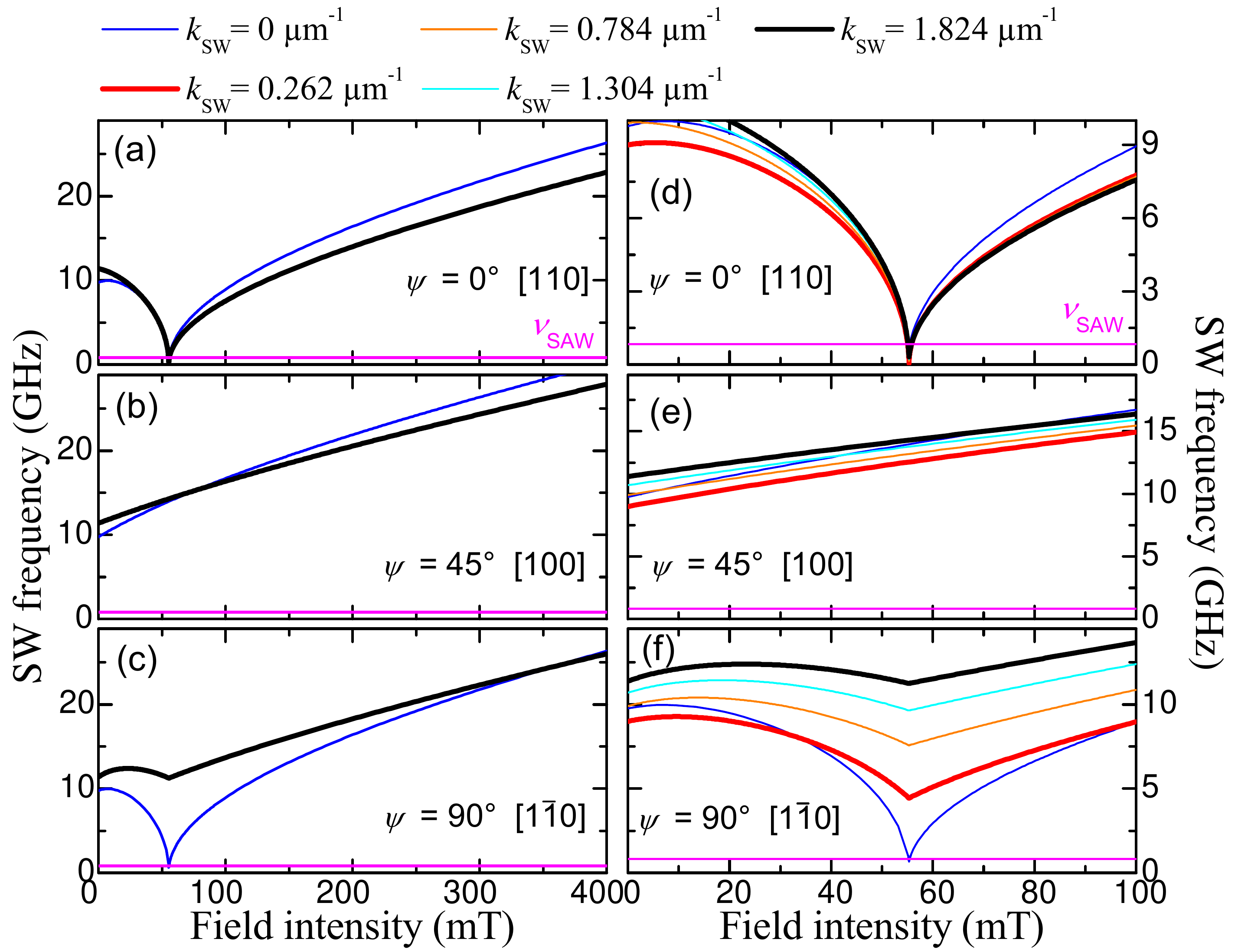}
    \caption{(a-c) : SW frequencies versus applied field magnitude, calculated for $k_{\rm SW}=0$ and for $k_{\rm SW}=k_{\rm SAW}=$1.824 $\mu$m$^{-1}$ in a 54-nm thick Fe film magnetized in plane, for three different orientations of the external field $\bf{B}_{\rm ext}$ applied at angles $\psi = 0^\circ, 45^\circ$, and $90^\circ$ from $\bf{k}_{\rm SAW}//[110]$. 
    (d-f): Zoom centered around $B_{\rm r} = 58$ mT of the SW frequencies, calculated for $k_{\rm SW}=0$ and for $k_{\rm SW} = k_{\rm SAW}$, where $k_{\rm SAW} =0.262$, 0.784, 1.304 and $1.824 ~\mu $m$^{-1}$ are the wavevectors of four SAWs with frequencies 119, 357, 595, and 833 MHz, respectively.
    The magenta line indicates the SAW frequency excitation at $\nu_{\mathrm{SAW}} = 833$ MHz}
    \label{fig:SW_disper}
\end{figure}

Before discussing the experimental results, it is useful to analyze the dependence of the lowest frequency SW mode as a function of the intensity of the magnetic field  $\mathbf{B}_{\rm ext}$ applied along different in-plane directions for a 54-nm thick Fe film. 
Figure~\ref{fig:SW_disper}~(a-c) reports the SWs frequency calculated at $k_{\rm SW}=0$ and $k_{\rm SW}$ = $k_{\rm SAW}$ = $1.824~\mu$m$^{-1}$ (i.e. the $\mathbf{k}$-vector of a SAW with frequency  $\nu_{\rm SAW}=833$ MHz), by means of the model described in Section \ref{sec:theo}, and using  the magnetic parameters obtained from the analysis of the Broad-Band-FMR (BB-FMR) and BLS measurements reported in Appendix B
\footnote{It turns out that the lowest frequency mode corresponds to the uniform mode for $k_{\rm SW}=0$ and to the Damon-Eshbach for $k_{\rm SW}$ different from zero.}.
Calculations were performed for $\mathbf{B}_{\rm ext}$ applied parallel to $[110]$, $[100]$ and $[1\bar{1}0]$ directions, at an angle $\psi$=0$^\circ$, 45$^\circ$ and 90$^\circ$ from $\mathbf{k}_{\rm SAW}$, respectively.
 
Note that $\psi$=0$^\circ$ and $\psi$=90$^\circ$ correspond to the Backward (BA) and Damon-Eshbach (DE) geometry, where  $\mathbf{k}_{\rm SW}$ is parallel and perpendicular to the sample magnetization, respectively. 
Figure~\ref{fig:SW_disper} (d-f) report SW frequencies calculated in a restricted range of the $\mathbf{B}_{\rm ext}$ intensity  for  $k_{\rm SW}=0$ and for $k_{\rm SW}$
equal to the $k_{\rm SAW}$ of  the four harmonic frequencies $\nu_{\rm SAW}$ used in the experiment. 
These calculations indicate that a wide spectrum of SW excitations  can be covered by a simple variation of the $\psi$-angle.

When the in-plane magnetic field is applied along the $[110]$ and $[1 \bar{1} 0]$  directions, corresponding to $\psi=0^{\circ}$ and $\psi=90^{\circ}$, respectively, the SW frequency exhibits a non-monotonic behavior, with a local frequency minimum at an external ﬁeld of about $B_r \approx 58$~mT, see Fig.~\ref{fig:SW_disper}~(a,c). This is the typical hard-axis behavior, and indicates a reorientation of the sample magnetization. On reducing the strength of the external ﬁeld, the magnetization does not remain oriented along the hard direction and starts to rotate towards the nearest easy axis, i.e. the in-plane $[100]$ directions, causing the increase of the SW frequency observed for field values smaller than $B_r$.  Moreover, one can note that in BA configuration the frequency dependence exhibits a marked minimum for all  $k_{\rm SAW}$  values, while in the DE configuration the frequency minimum becomes less pronounced on increasing the  $k_{\rm SAW}$ magnitude, see Fig.~\ref{fig:SW_disper}~(d,f).

In contrast, when $\mathbf{B}_{\rm ext}$ is parallel to the easy axis $[100]$ direction (corresponding to $\psi=45^{\circ}$), the SW frequency shows a monotonic dependence as a function of ${B}_{\rm ext}$, see Fig.~\ref{fig:SW_disper}~(b,e). 
Since  in our experiment $\nu_{\rm SAW}$ is smaller than 1 GHz, these calculations indicate that the resonant magnetoelastic coupling effect (res-MEC), where both the frequencies and the $\mathbf{k}$-vectors of SAW and SW match, can be obtained only in the BA geometry. However, as shown previously in Ref.~\onlinecite{Duquesne2019a}, a slight misalignment of just $0.1^\circ$ in $\psi$ is capable to drive the system off-resonance because of the rapid increase in the precession frequencies.

\section {Results and discussion}

\subsection{Comparison between different magnetic field directions}

\begin{figure}
    \centering
    \includegraphics[width=\linewidth]{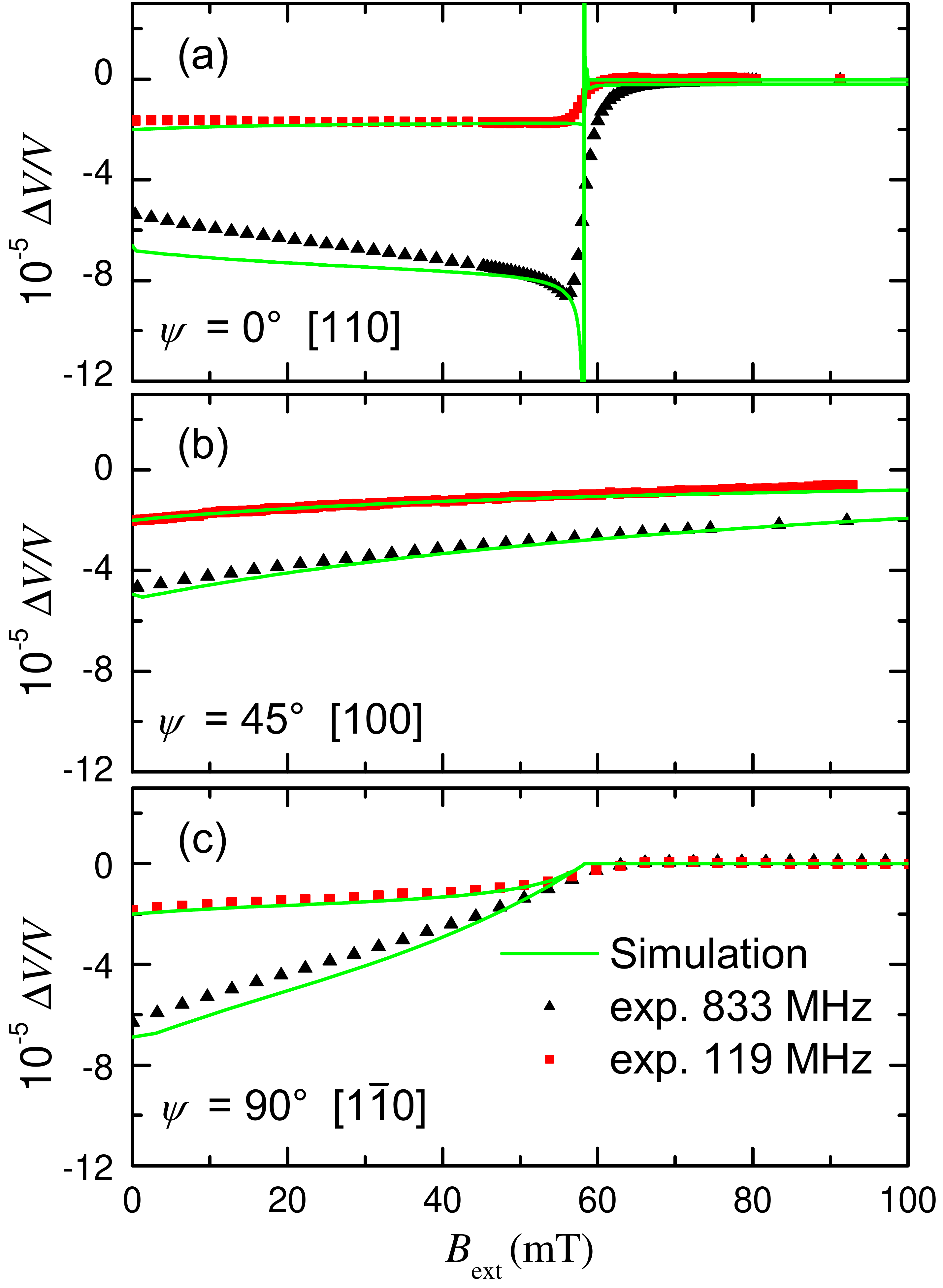}
    \caption{Relative variation of the SAW velocity $V$ measured versus decreasing amplitude of the external field $B_{\rm ext}$ at $\nu_{\rm SAW}=119$ MHz and 833~MHz. The magnetic field $\mathbf{B}_{\rm ext}$ is applied in the film plane, either parallel (a) to the $[110]$ direction (hard axis, $\psi = 0^\circ$), or (b) to the $[100]$ direction (easy axis, $\psi = 45^\circ$), or (c) to the $[1\bar{1}0]$ direction (hard axis, $\psi = 90^\circ$). 
    At $\psi = 0^\circ$, $\mathbf{k}_{\rm SAW}$ and $\mathbf{B}_{\rm ext}$ are parallel, with $\mathbf{k}_{\rm SAW}$ parallel to the $[110]$ direction. The lines correspond to the simulation results using the model in subsection \ref{sec:theo} i.e. taking into account both the Rayleigh wave and the correct SWs dispersion with a Gilbert constant $\alpha = 0.005$. 
    }
    \label{fig:exp2}
\end{figure}

\begin{figure}
    \centering
    \includegraphics[width=\linewidth]{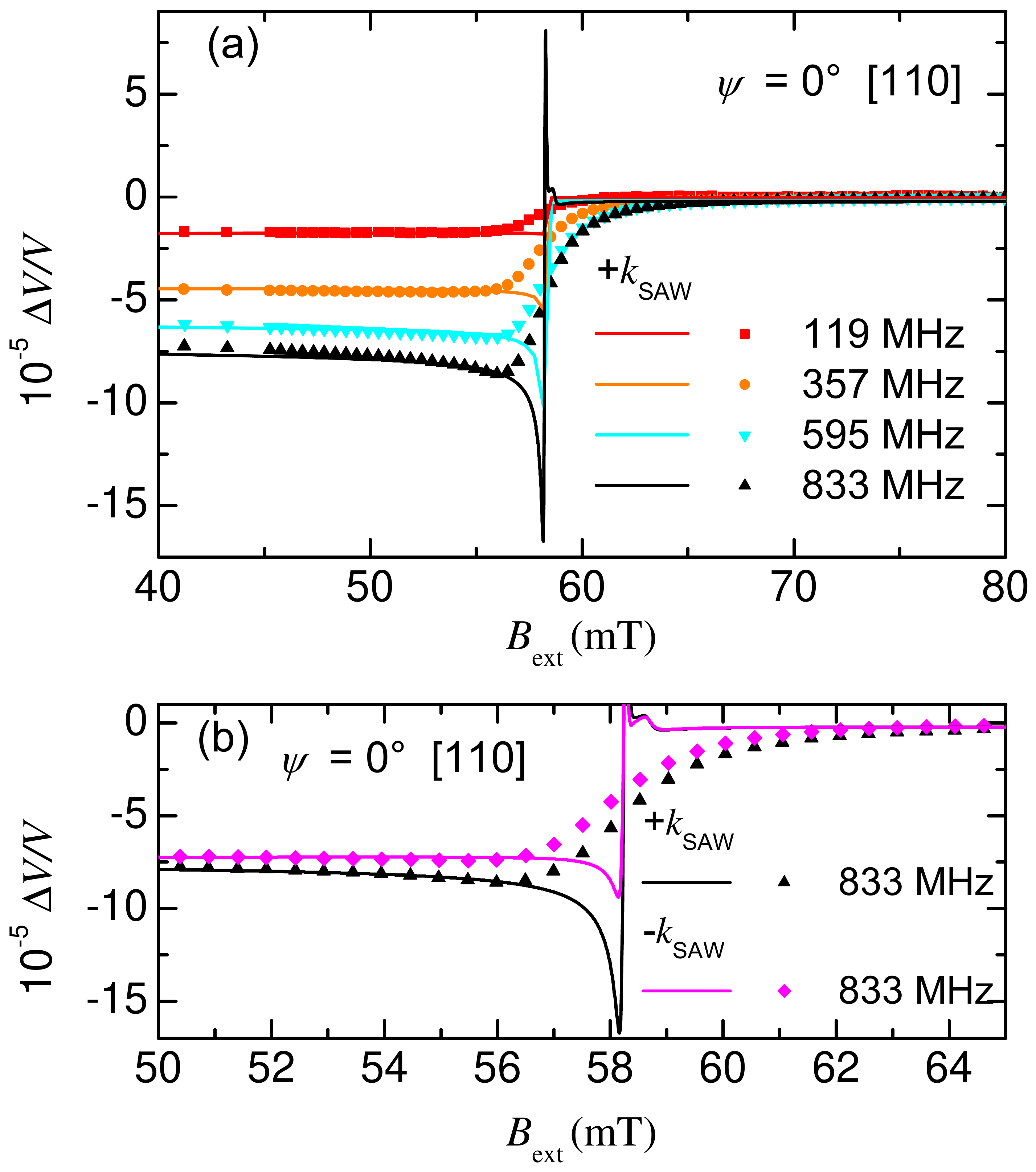}
    \caption{(a) Relative variation of the SAW velocity $V$ measured versus decreasing amplitude of the external field $\mathbf{B}_{\rm ext}$~//~$[110]$ (i.e., $\psi =0^{\circ}$) at $\nu_{\rm SAW}=119$, 357, 595, and 833~MHz.  
    $\mathbf{k}_{\rm SAW}$ and $\mathbf{B}_{\rm ext}$ are parallel, with $\mathbf{k}_{\rm SAW}$ parallel to the $[110]$ direction. The solid lines correspond to the simulation results using the model in subsection \ref{sec:theo}. (b) At 833 MHz, both $\mathbf{+ k}_{\rm SAW}$ and $\mathbf{- k}_{\rm SAW}$ experimental (symbols) and simulation curves (solid lines) are shown and correspond to SAW propagation direction parallel and antiparallel to [110], respectively.
    }
    \label{fig:exp}
\end{figure}

Figure \ref{fig:exp2} (a-c), report  the velocity changes measured at  $\nu_{\rm SAW}$~=~119~MHz and~833~MHz, in three different configurations: (a) $\mathbf{B}_{\rm ext}$~//~$[110]$ ($\psi =0^{\circ}$), (b) $\mathbf{B}_{\rm ext}$~//~$[100]$ ($\psi =45^{\circ}$), and (c) $\mathbf{B}_{\rm ext}$~//~$[1\bar{1}0]$ ($\psi =90^{\circ}$). 
Figure \ref{fig:exp} shows more detailed measurements performed at $\psi =0^{\circ}$,  at the four harmonic frequencies $\nu_{\rm SAW}$ (panel (a)) and applying the external magnetic field both parallel and antiparallel to the $\mathbf{k}_{\rm SAW}$ direction (panel (b)).
It’s worth noticing that the velocity change strongly depends on the angle between the direction of $\mathbf{B}_{\rm ext}$ and $k_{\rm SAW}$.  When $\mathbf{B}_{\rm ext}$~//~$[110]$  the relative variation of SAW velocity presents a sharp change at the saturation magnetic field, $B_r$, which becomes less pronounced on reducing the SAW frequency. This behaviour is a signature of the res-MEC conditions  as demonstrated in Ref.~\onlinecite{Duquesne2019a}. In this geometry, indeed, the resonance conditions are fulfilled owing to the dramatic lowering of the  SWs  frequencies at $B_r$ , as discussed above and presented in Fig.~\ref{fig:SW_disper}~(a, d).

On the contrary, a smoother change is observed for the velocity variation when $\mathbf{B}_{\rm ext}$~//~$[1\bar{1}0]$ (Fig.~\ref{fig:exp2}~(c)) at both 833 and 119 MHz. In this configuration, the frequency matching, i.e. $\nu_{\rm SAW}$~=~$\nu_{\rm SW}$, is obtained only at  $k_{\rm SW}$=0, while the matching is rapidly lost when  $k_{\rm SW}$ increases, as it can be seen in  Fig.~\ref{fig:SW_disper}~(c,f). This behaviour can be explained taking into account the large dispersion of the lowest frequency mode in DE geometry (see Fig.~\ref{fig:B2} in the Appendix B). Therefore, at $\psi = 90^{\circ}$ the measured velocity variation comes from a non-res-MEC, and can be ascribed to the in-plane rotation of the sample magnetization towards the easy axis, when the intensity of the magnetic field is reduced. \\
Eventually, a constant decrease of the velocity variation  is observed on decreasing the field amplitude when $\mathbf{B}_{\rm ext}$~//~$[100]$ (Fig.~\ref{fig:exp2}~(b)) at both 833 and 119 MHz, since  at $\psi = 45^\circ$ the resonant condition $\nu_{\rm SAW}$ = $\nu_{\rm SW}$ is far from being satisfied, as one can see in Fig.~\ref{fig:SW_disper}~(b, e).

Finally, as it can be seen in Fig.~\ref{fig:exp}~(b) a non-reciprocal SAW propagation is observed when $\mathbf{B}_{\rm ext}$~//~[110], whereas the SAW non-reciprocity is not found when $\mathbf{B}_{\rm ext}$~//~$[1\bar{1}0]$ and $\mathbf{B}_{\rm ext}$~//~$[100]$. 
We recall that non-reciprocity means that SAW propagation (attenuation and velocity) depends on whether the acoustic wavevector is parallel or antiparallel to a given direction \cite{Sasaki2017}.

It is important to note that in Ref.~\onlinecite{Duquesne2019a}, SAW attenuation was found to change as a function of the magnetic field intensity only  for $\psi = 0^{\circ}$,  while for $\psi = 45^{\circ}$ and $90^{\circ}$, the variation of SAW attenuation was negligible. Here, changes of the relative velocity variation were observed in all the investigate geometries and even off-resonance, suggesting that the relative velocity is much more sensitive to the magnetic configuration than the acoustic attenuation.

In the next paragraph we will describe the phenomenological model used to explain these experimental results.\\

\subsection{Phenomenological magnetoelastic approach considering spin wave dispersion} 
\label{sec:theo}

In this section we present the model used to interpret the evolution of the velocity variation as a function of both the in-plane direction and the intensity of the applied magnetic field. In a previous work Dreher \textit{et al.} \cite{Dreher2012} developed a phenomenological approach to explain the dependence of the SAW attenuation as a function of the in-plane and out-of-plane angle between the applied magnetic field and the SAW propagation direction. In this approach the authors solved the LLG equation to obtain expressions for the magnetization dynamics and then took into account a purely longitudinal acoustic wave, instead of the more complicated Rayleigh wave, to establish the back-action of the ferromagnetic resonance on  the acoustic wave.  Subsequently both the longitudinal and the transverse terms of the Rayleigh wave were successfully taken into account in a semi-infinite medium approximation \cite{Thevenard2014} and in a layered system by Hern\'andez \textit{et al.} \cite{Hernandez-Minguez2020,Kuszewski_2018} where the authors succeeded in describing the non-reciprocity of SAW attenuation.
Previous studies by other authors have already shown the importance of dispersion of spin waves in attenuation analysis \cite{Hernandez-Minguez2020, Gowtham2015}. Here, we show that it's essential to take into account the dispersion of spin waves to explain the behavior of the velocity as a function of the applied field.
Our starting point, in order to evaluate the relative SAW velocity change $\Delta V/V$ in epitaxial Fe thin films on a GaAs(001) substrate, is the following \textit{ansatz} which we adopt for Rayleigh waves \cite{Dreher2012}  \footnote{This equation is correlated with the quasi-exact model in Section \ref{sec:quasi} via the evaluation of a proportionality factor R.}:
\begin{equation}
    R_{\rm eff} \Delta P = -2 x_0 \Delta \widetilde{k}_{\rm SAW} P_{ac}^{layer}
    \label{eqP}
\end{equation}

$\Delta P$ is the maximum power that can be transmitted from the SAW to the magnetization dynamics (i.e. to SWs) and $R_{\rm eff}$ is an \textit{ad hoc} parameter to compare calculations with experimental results.
$P_{ac}^{layer}$ is the acoustic power carried by travelling SAWs in the layer (note that, contrary to Dreher \textit{et al.} \cite{Dreher2012} who took into account the acoustic power of the wave we consider here only the acoustic power in the layer); $x_0$, the propagation length of the magnetic device ($x_0 = 4$~mm in our case); and $\Delta \widetilde{k}_{\rm SAW}$, a small perturbation of the wave number of the SAW. 
We assume a vector displacement $\mathbf{u}\mathrm{(X)} = \mathbf{u_0} e^{i(\widetilde{k}X -\omega t)}$ where $\widetilde{k}_{\rm SAW} = k_{\rm SAW}+iA$, and the amplitude $A$ (in m$^{-1}$) is linked to the attenuation $\Gamma$ (in $\rm{dB} \cdot  \rm{m}^{-1}$) by the expression $\Gamma = (20/\ln 10) A$. Consequently, the complex quantity $\Delta \widetilde{k}_{\rm SAW}$ is related  to the variation of the SAW velocity ($\Delta V$) and of the attenuation ($\Delta \Gamma$) by:
\begin{equation}
    \frac{\Delta V}{V} = -  \frac{Re (\Delta \widetilde{k}_{\rm SAW})}{k_{\rm SAW}} , ~~~~~
    \Delta \Gamma = \frac{20}{\ln10} Im ( \Delta \widetilde{k}_{\rm SAW} )
\end{equation}

The acoustic power can be derived from the approach given by Royer in Ref. \onlinecite{Royer_book}: 
\begin{equation}
    P_{ac}^{layer} = \frac{1}{2} \rho w v_R \omega^2 \int_{0}^{d} \left [ |u_{X}|^2 + |u_{Y}|^2 + |u_{Z}|^2 \right ] \,dz 
    \label{eq3}
\end{equation}
where the $u_i$ components are defined in the rotated frame $(X // [110], Y //[1\overline{1}0], Z //[001])$, $w = 4$~mm is the acoustic beam width, $v_{R} $ is the Rayleigh sound velocity in the $[110]$ direction (see Tab.~\ref{tab:epsilon}), $\rho = 7851$~kg.m$^{-3}$ is the Fe mass density \cite{Adams2006} and $d$ is the magnetic film thickness (54 nm). In our case, $u_{Y} = 0$.\\

The electromagnetic power, $\Delta P$, transmitted to the magnetic layer is calculated as follows \cite{Dreher2012,Gowtham2015,Hernandez-Minguez2020}:
\begin{equation}\label{DeltaP}
    \Delta P = - \frac{\omega \mu_0 M_s}{2} \int_{V_0} \left [ (h_\theta^* , h_\varphi^*) \bar{\chi} \begin{pmatrix}
h_\theta \\
h_\varphi 
\end{pmatrix}  \right ] dV 
\end{equation}

where $\theta$ and $\varphi$ denote, respectively, the polar and the azimuthal (with respect to the $[110]$ direction) angle of the normalized magnetization vector, i.e. $\textbf{m}=\textbf{M}/M_s$ where $M_s$ is the saturation magnetization and $V_0$ is the Fe thin film volume. $\bar{\chi}$ is the Polder susceptibility matrix \cite{Hernandez-Minguez2020} which describes the magnetic response of the ferromagnet to a small, time-varying  magnetoelastic field (associated with the travelling SAW) whose out-of-plane ($h_{\theta}$) and in-plane ($h_{\varphi}$) components are perpendicularly oriented with respect to the equilibrium magnetization $\textbf{m}_0$. The explicit forms for $h_{\theta}$, $h_{\varphi}$ and $\bar{\chi}$ in Eq. (\ref{DeltaP}) can be obtained \cite{Hernandez-Minguez2020} in the framework of the Landau-Lifshitz-Gilbert \cite{LLGa} equation of motion for $\textbf{m}$:
\begin{equation}
\partial_t \textbf{m}= - \gamma \textbf{m} \times \mu_0 \textbf{H}_{\rm eff} + \alpha \textbf{m} \times \partial_t \textbf{m}
\end{equation}
where $\gamma>0$ is the gyromagnetic ratio and $\alpha$ is a phenomenological damping parameter \cite{LLGa}. The time dependence of $\textbf{m}$ is determined by the effective magnetic field $\mu_0 \textbf{H}_{\rm eff}=-\nabla_{\textbf{m}} f_{tot}$, where $f_{tot}=f+f_{me}+f_{rot}$ is the total free energy density (normalized to the saturation magnetization ${M_s}$), consisting of a purely magnetic contribution, $f$ (reported in Appendix A), a magnetoelastic one, $f_{me}$, and a magneto-rotation one, $f_{rot}$.
For a cubic solid, $f_{me}$ reads:

\begin{equation}\label{fme}
\begin{split}
f_{me} &= B_1 \Big[  \varepsilon_{xx}\Big( m_x^2-\frac{1}{3} \Big) + \varepsilon_{yy}\Big( m_y^2-\frac{1}{3} \Big)\\
&+ \varepsilon_{zz}\Big( m_z^2-\frac{1}{3} \Big) \Big] \\
&+ 2 B_2 ( \varepsilon_{xy} m_x m_y +\varepsilon_{yz} m_y m_z  + \varepsilon_{zx} m_z m_x )
\end{split}
\end{equation}
where $B_1$ and $B_2$ are phenomenological magnetoelastic coupling constants \cite{Duquesne2019a}, $\varepsilon_{ij}$ are strain components ($i,j=x,y,z$) expressed in the standard cubic frame ($x //[100], y // [010], z // [001]$), and $m_i$ is the normalized magnetization component along $i$.
mostly due to the shape anisotropy field directed along the $z$ axis. Similarly to Xu \textit{et al.} \cite{Xu2020}, we adopt the energy term reported by Tremolet in Ref. \onlinecite{Tremolet}~p.~78 for a tetragonal symmetry by considering that the magnetic shape anisotropy term is one order of magnitude larger that the magnetocrystalline out-of-plane uniaxial anisotropy, $K_u$ (see Appendix B). As a consequence, the energy reads as follows:

\begin{equation}
    f_{rot} = \mu_0 M_S^2 (\omega_{yz} m_y m_z + \omega_{xz} m_x m_z)     
\end{equation}
where 
$\omega_{ij} =\dfrac{1}{2}(\dfrac{\partial u_{i}}{\partial x_j} - \dfrac{\partial u_{j}}{\partial x_i})$  are the components of the rotation tensor \footnote{The cubic symmetry for the magnetoelastic energy of Eq.~\ref{fme} still holds despite the overall tetragonal symmetry as shown in Barturen \textit{et al.}\cite{Barturen} .}.

Linearizing the LLG equation of motion, i.e. considering small deviations,  $\delta \theta$ and $\delta \varphi$, in the two directions perpendicular to the equilibrium magnetization $\textbf{m}_0$, one obtains \cite{Hernandez-Minguez2020}: 
\begin{equation}\label{deltachidelta}
\begin{pmatrix}
\delta\theta \\
\delta\varphi 
\end{pmatrix} = \bar{\chi} \begin{pmatrix}
h_\theta \\
h_\varphi 
\end{pmatrix}
\end{equation}
For in-plane $\textbf{m}_0$ (i.e. $\theta_0 = \pi/2$, and $\varphi_0$  given by Eq. \ref{equilibrium} in Appendix A), the effective field is obtain by calculating  $-\nabla_{\textbf{m}}( f_{me} + f_{rot})$. Thus, the out-of-plane and the in-plane components of the effective field read as follows \cite{Hernandez-Minguez2020, Xu2020}:

\begin{subequations}
\begin{align}
        \mu_0 h_\theta &= \frac{2B_2}{M_S} \cos{\varphi_0}~ \varepsilon_{XZ} -\mu_0 M_s \cos{\varphi_0}~ \omega_{XZ} \label{muhtheta}\\
       \mu_0 h_\varphi &= \frac{2B_2}{M_S} \sin{\varphi_0} \cos{\varphi_0}~ \varepsilon_{XX} \label{muhphi}
\end{align}
\end{subequations}

\noindent where $\varepsilon_{ij}$ ($i,j=X,Y,Z$) are the dynamical strain components defined in the rotated frame. 
\textit{It is very important to report that, we could have limited our approach to the in-plane field (Eq.~\ref{muhphi}) to give a satisfying description of the observed velocity field dependence. We decided to include the $h_\theta$ field into our calculations for completeness and accuracy and to estimate the expected non reciprocity effects} \footnote{$\omega_{XZ}$ is much smaller than $\varepsilon_{XZ}$ at the surface (see Tab.~\ref{tab:epsilon}). At odds with Refs. \cite{Hernandez-Minguez2020} and \cite{Sasaki2017}, we consider that the two terms of Eq.~\ref{muhtheta} must be retained, as shown by \cite{Xu2020}}.

The Polder matrix $\bar{\chi}$ takes the form \cite{Hernandez-Minguez2020} \footnote{The matrix was not inverted and the saturated magnetization $M_s$ was missing.}:
\begin{equation}\label{pmatrix}
    \bar{\chi} = \frac{\gamma \mu_0}{D}
    \begin{pmatrix}
        \gamma f_{\varphi\varphi}(\textbf{k}
       _{\rm SW}
        )-i\alpha\omega & -i \omega \\
        i\omega & \gamma f_{\theta\theta}(\textbf{k}
       _{\rm SW}
        )-i\alpha \omega
    \end{pmatrix}
\end{equation}
with  
\begin{align}
    D & = -(1+\alpha^2)\Big(\omega^2-\omega_0^2(\textbf{k}
    _{\rm SW}
    )-i\omega\kappa\Big) \\
    \omega_{0}^2 (\textbf{k}
    _{\rm SW}
    )& = \frac{\gamma^2}{\alpha^2+1} f_{\varphi\varphi}(\textbf{k}
    _{\rm SW}
    )f_{\theta\theta} (\textbf{k}
    _{\rm SW}
    )\\
    \kappa & = \gamma \frac{\alpha}{\alpha^2+1} \Big(f_{\varphi\varphi}(\textbf{k}
    _{\rm SW}
    )+f_{\theta\theta}(\textbf{k}
    _{\rm SW}
    )\Big)
\end{align}

It is fundamental to notice the important ingredient of our approach: in our calculation of the Polder susceptibility, the spin-wave wavevector $\textbf{k}_{\rm SW}$ is explicitly taken into account through the dispersion relation of the lowest-frequency spin-wave mode, $\omega_{0} (\textbf{k}_{\rm SW})$, and the second derivatives of the purely magnetic free energy density, $f$.
 
Using dipole-exchange spin-wave theory \cite{KS1986} for a tangentially magnetized ferromagnetic film with thickness $d$, and neglecting the dipolar-induced hybridization between spin-wave modes in the low wavevector limit ($kd <<1$), the following approximate analytical expressions  for $f_{\theta\theta}(\textbf{k}_{\rm SW})$ and $f_{\varphi\varphi}(\textbf{k}_{\rm SW})$ are obtained \cite{Tacchi2019}
\begin{equation}\label{ftt}
\begin{split}
&f_{\theta\theta}(\textbf{k}_{\rm SW})=\mu_0 H \cos{(\varphi_0-\psi)}+\mu_0 M_s(1-P_{00})\\
&+\mu_0 H_1 ~ \frac{1}{4}\left[3-\cos{(4\varphi_0)}\right]-\mu_0 H_u
+\mu_0 H_{ex} (ka)^2
\end{split}
\end{equation}
\begin{equation}\label{fff}
\begin{split}
&f_{\varphi\varphi}(\textbf{k}_{\rm SW})=\mu_0 H \cos{(\varphi_0-\psi)}+\mu_0 M_s P_{00} \sin^2{\varphi_0}\\
&-\mu_0 H_1\cos{(4\varphi_0)}+\mu_0 H_{ex} (k a)^2
\end{split}
\end{equation}
where $\mu_0 H$ is the intensity of the external magnetic field; $\mu_0H_1={2K_1}/{M_s}$, $\mu_0H_u={2K_u}/{M_s}$, and $\mu_0H_{ex}={(2A_{ex})}/{(M_s a^2)}$ denote effective magnetic field intensities, respectively associated with the cubic anisotropy energy density ($K_1$), the out-of-plane uniaxial anisotropy energy density ($K_u$), and the exchange energy density ($A_{ex}/a^2$, where $a$ is the cubic cell length of Fe); $\mu_0 M_s$ is a magnetic dipolar field, and $P_{00}=1-[\frac{1-e^{-kd}}{kd}]$ a dimensionless dipolar factor. 

It is now worth noticing that, in the limit of very low wavevector ($kd <<1$), analytical expressions \cite{Tacchi2019} (not reported here) can be obtained  also for the higher-frequency spin wave modes (i.e., $\omega_{n}(\bf{k}_{\rm SW})$ with $n=1,2,\cdots$), where the spin-wave wavevector $\bf{k}_{\rm SW}$ can be approximately separated in a component ($\textbf{k}$) tangential to the film plane, and the other ($k_{\perp,n}=n{\pi}/{d}$ with $n=1,2,\cdots$) perpendicular to the film plane. In the general case of finite in-plane wavevector, the calculation of the dipole-exchange spin-wave frequencies  $\omega_{n}(\bf{k}_{\rm SW})$ ($n=0,1,2,\cdots$) requires, instead, the numerical diagonalization of a square matrix \cite{KS1986,Tacchi2019} because magnetic dipole-dipole interactions provide substantial hybridization between the SW modes for $kd \gtrsim 1$ \cite{KS1986,Tacchi2019}.

In Appendix B we made a comparison (see Figs. \ref{fig:B0}--\ref{fig:B2}) between BLS measurements of the SW frequencies and theoretical expressions for the three lowest SW modes ($n=0,1,2$). In this way, we could obtain a quantitative evaluation of both the magnetic parameters and the thickness of our Fe film.  

Finally, it is fundamental to remark  that, on the basis of the SW dispersion relation measured by BLS for k $< 0.2\times 10^7$ rad/m (very low wavevector, see Fig.~\ref{fig:B2} of the Appendix B), the lowest frequency mode corresponds to the mode $n=0$. Therefore, in our calculation of the Polder susceptibility (Eq.~\ref{pmatrix}), only the lowest-frequency spin-wave mode, $\omega_{0} (\textbf{k}_{\rm SW})$, was taken into account.

\subsection{Comparison with experimental results}

Figures \ref{fig:exp2} and \ref{fig:exp} report the comparison between the experimental and the calculated $\mathbf{B}_{\rm ext}$-dependence of the velocity changes. Theoretical curves were numerically obtained using the approach described in the previous paragraph.   

As it can be seen, a very good agreement is found for different directions of the magnetic field (Fig.~\ref{fig:exp2}). In addition for $\psi =0^{\circ}$ the theoretical calculations reproduce quite well the evolution of $\Delta V/V$ as a function of $\nu_{\rm SAW}$ for all frequencies (Fig.~\ref{fig:exp}) showing that the factor $R_{\rm eff}$ is proportional to the SAW frequency. In particular, it turns out that the best-fit proportionality factor is $R_{\rm eff} = \nu_{\rm SAW}/7050$ ($\nu_{\rm SAW}$ in MHz)  permitting us to infer that this approach should be valid even at higher frequencies, provided that $\nu_{\rm SAW} < 7~050$ MHz.

It's important to report that similar calculations performed in the uniform ($k_{\rm SAW}=0$) approximation give very similar trends for the $\psi=0^\circ$ and $\psi=90^\circ$ geometries, with sharp variations at $B_r$. However, experimentally these two configurations do not show the same trend. Consequently, the k-dispersion has a strong impact on magnetic field induced velocity changes. 

Note that the theoretical curves calculated for $\mathbf{B}_{\rm ext}$ // [110] present a much sharper variation than the experimental ones. We ascribe this observation to the mosaicity of the sample [the FWHM of the (002) planes is approximately 0.3$^\circ$, measured by x-ray diffraction], which is responsible for a spreading of the resonant frequencies at a given $B_{\rm ext}$ \cite{Duquesne2019a}.

In the previous chapter we anticipated that a satisfying description of the observed velocity change field dependence is obtained by considering that $h_\theta$ =0 in Eq.~\ref{muhtheta}. This is described in Appendix C where a good agreement between experimental results and a fully longitudinal wave approach is found for $R_{\rm eff} = \nu_{\rm SAW}/24000$ ($\nu_{\rm SAW}$ in MHz). This approach is much simpler as already shown for SAW attenuation measurements in Ref. \cite{Dreher2012}. Interestingly, the $h_\theta$ component is crucial to understand the origin of non-reciprocal phenomena. Indeed, following the Xu \textit{et al.} arguments, we find out that the magneto-rotational coupling, i.e. the second term of the $h_\theta$ field in Eq.~\ref{muhtheta}, is the source of the observed non-reciprocity in Fig.~\ref{fig:exp} (b) \cite{Xu2020}. Indeed, when this term is not taken into account a nearly perfect reciprocity is recovered (not shown). 

\begin{figure}
    \centering
    \includegraphics[width=\linewidth]{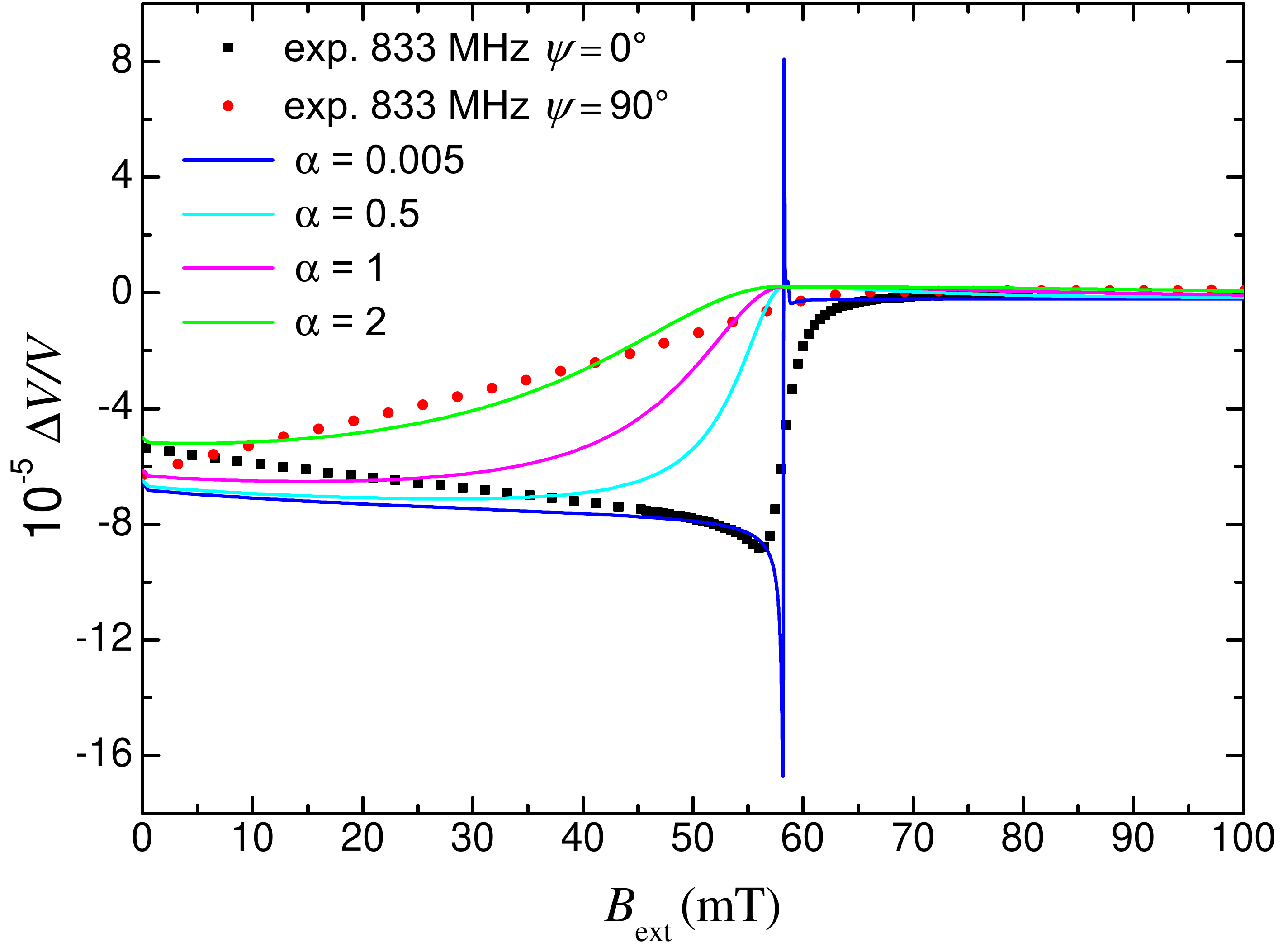}
    \caption{Relative variation of the SAW velocity $V$ measured versus decreasing amplitude of the external field $B_{\rm ext}$ at $\nu_{\rm SAW}=833$~MHz at $\psi = 0^\circ$ and $\psi = 90^\circ$ as represented in Fig.~\ref{fig:exp2}~(a, c). The solid lines correspond to the simulation results using the model in subsection \ref{sec:theo} for different values of the Gilbert parameter at $\psi = 0^\circ$. The $\alpha$ parameter was exaggerated in order to let the resonant dynamical effects fade out.}
    \label{fig:damping}
\end{figure}

Finally, in order to put in evidence the impact of  res-MEC and of magnetization dynamics on the velocity variation, we calculate  $\Delta V/V$ on changing the Gilbert damping, $\alpha$, (Fig.~\ref{fig:damping}). One can see that the sharp velocity change observed around $B_r~\approx~58$~mT  for $\psi=0^\circ$ smears out when the dynamics is turned
off by an artiﬁcially high value of $\alpha$.
We notice that the over-damped velocity change obtained with $\alpha = 2$ well describes the experimental results for $\mathbf{B}_{\rm ext}$ $\perp$ $\mathbf{k}_{\rm SAW}$ ($\psi=90^\circ$). This permits us to conclude that the sharp-shape velocity change observed for $\psi=0^\circ$ is a genuine fingerprint of the resonant SAW-induced magnetization precession, and that the smoother $\mathbf{B}_{\rm ext}$ dependence observed in the case $\psi=90^\circ$ is the non-resonant expected trend.

\subsection{ Fully magnetoelastic approach for k$_{\mathrm{SAW}}$~=~0} 
\label{sec:quasi}

In order to corroborate our phenomenological approach based on Eq. \ref{eqP}, we developed an alternative fully elastic approach based on SAW propagation in a layered structure. Here, we consider a thin Fe film epitaxied on top of a GaAs substrate. Both systems present a cubic structure with the same orientation. Magnetism is considered by introducing additional magnetic terms in the stress versus strain relations. In case of in-plane equilibrium magnetization ($\theta_0=\pi/2$ and $\varphi=\varphi_0$), we read \cite{hepburn:tel-02411563}:
\begin{equation}
\label{eq:contraintes_couche}
\left\{
\begin{array}{lll}
  \sigma_{XX} & = &
  \widehat{C'}_{11} \varepsilon_{XX} +
  \widehat{C'}_{13} \varepsilon_{ZZ}
  +
  B_2 \cos (2{\varphi_0}) ~ \delta \varphi 
  \\
  \sigma_{YY} & = &
  \widehat{C'}_{12} \varepsilon_{XX} +
  \widehat{C'}_{23}\varepsilon_{ZZ}
  -
  B_2 \cos (2{\varphi_0}) ~ \delta \varphi 
  \\
  \sigma_{ZZ} & = &
  \widehat{C'}_{13} \varepsilon_{XX} +  \widehat{C'}_{33} \varepsilon_{ZZ}
   \\
  \sigma_{XY}    & = &  B_{1} \sin (2{\varphi_0})  ~ \delta \varphi
  \\
  \sigma_{XZ}    & = & 2 \widehat{C'}_{55}  \varepsilon_{XZ} -  B_{2} \sin ({\varphi_0} + \pi/4 ) ~ \delta \theta			  
  \\
  \sigma_{YZ}    & = &  -  B_{2} \sin ({\varphi_0} - \pi/4 ) ~ \delta \theta
  \end{array}
\right.
\end{equation}

\noindent where 
$B_{1}$ and $B_{2}$ are magnetoelastic constants, defined in the standard cubic frame, i.e. axis parallel to $[100], [010], [001]$ directions (see Eq.~\ref{fme}). $\sigma_{ij}$, $\varepsilon_{ij}$ and $\widehat{C'}_{ij}$ are the stress, strain and elastic constants, in the layer, written in a rotated frame, i.e. axis parallel to $[110], [\bar{1}10], [001]$ directions (elastic constants values in Appendix D).

In both layer and substrate, we assume that the SAW is a linear combination of partial waves. This is the standard procedure in the derivation of the surface waves in purely elastic layered structure. Partial waves read
\begin{equation}
\label{eq:partielle_substrat}
\left\{
\begin{array}{lll}
   u_{X} & = & U~\exp(-\xi Z) \exp i(k X - \omega t) \\
   u_{Y} & = & 0 \\
   u_{Z} & = & W~\exp(-\xi Z) \exp i(k X - \omega t)
\end{array}
\right.
\end{equation}
in the substrate, and
\begin{equation}
\label{eq:partielle_couche}
\left\{
\begin{array}{lll}
   \widehat{u}_{X} & = & \widehat{U}~\exp(-\widehat{\xi} Z) \exp i( k X - \omega t) \\
   \widehat{u}_{Y} & = & 0 \\
   \widehat{u}_{Z} & = & \widehat{W}~\exp(-\widehat{\xi} Z) \exp i(k X - \omega t) \\
   \delta \varphi & = & \Phi ~\exp(-\widehat{\xi} Z) \exp i(k X - \omega t) 	\\
   \delta \theta & = & \Theta ~\exp(-\widehat{\xi} Z) \exp i(k X - \omega t) 
\end{array}
\right.
\end{equation}
in the layer, where $u_{i}$ and $\widehat{u}_{i}$ are the displacement fields in the rotated frame, in the layer and substrate, respectively. $X$ and $Z$ are the coordinates along the propagation axis $[110]$ and along the out-of-plane axis $[001]$, respectively.

We then look for a solution satisfying both the equations of motion and the elastic boundary conditions, at the surface and the interface. 
This calculation is quite tedious but it makes the bridge between theory of elasticity and magnetization dynamics. Since our aim is to corroborate the intensity of the $R_{\rm eff}$-factor and understand its origin, we simplify our approach i) by neglecting the magnetic boundary conditions as in Refs.~\onlinecite{Dreher2012} and \onlinecite{Hernandez-Minguez2020}, ii) by solving the LLG equations in a uniform mode approach and iii) by neglecting the magneto-rotational term. 
This approximation is valid for $\nu_{\rm SAW}$~=~119 MHz (corresponding to $k_{\rm SAW}~=~0.262$~$\mu$m$^{-1}$) and $\psi=0^\circ$ for which  the magnetic dynamic parameters $f_{\theta\theta}$ and $f_{\varphi\varphi}$ in the uniform mode approximation are very close to the exact values (not shown here). 
We also assume that the damping term $\alpha$ is zero. We also consider that non-reciprocity is a small effect that can be neglected for this elastic approach.
Using this approach we calculate $\Delta k_{\rm SAW}/k_{\rm SAW}$ and $\Delta P/P_{ac}^{layer}$ and then we obtain $R_{\rm eff} = \nu_{\rm SAW} / 9434$. We consider that the values of $R_{\rm eff}$ determined by the two models are in good agreement considering the accuracy of the models.

This result is important since it demonstrates that the intensity of the MEC interaction is fully described by a standard model based on elasticity, describing SAW propagation in a layered elastic structure: i.e., no hidden or more complicate interactions have to be considered. Moreover, we show that the much simpler phenomenological magnetoelastic approach based on Eq.~\ref{eqP} and presented in Section \ref{sec:theo} can be safely adopted to describe the magnetoelastic coupling in Fe thin films, once that $R_{\rm eff} \approx$ $\nu_{\rm SAW} / 7050$.

\section{Conclusions}

In this work we have investigated the dependence of the SAW velocity variation  as a function of the in-plane direction of the external applied magnetic field $\mathbf{B}_{\rm ext}$ in a Fe thin film grown on a GaAs  substrate.  In order to explain the experimental results we have exploited a phenomenological approach, where the SW dispersion has been explicitly included by solving the LLG equation. 
We have demonstrated the $\Delta V/V$ not only depend on the orientation between $\mathbf{B}_{\rm ext}$ and $\mathbf{k}_{\rm SAW}$, but it is also more sensitive to the SW dispersion than the SAW attenuation.  
Moreover, we have introduced a proportionality parameter $R_{\rm eff}$ that describes the intensity of the magnetoacoustic interaction of SAWs travelling in a multilayer, i.e. a magnetic thin film over a piezoelectric substrate. The parameter $R_{\rm eff}$ is observed to be proportional to the SAW frequency. Remarkably, the best-fitted value of this parameter $R_{\rm eff}= \nu_{\rm SAW} /7050$ is found to describe well the velocity change for all the probed frequencies and magnetic field directions. Despite the approximation involved, this approach gives a very good description of the SAW-SW interaction in the (\textbf{k},$\omega$) space. 
In this context, our findings give a comprehensive and quantitative picture of the magnetic field control of the SAW-SW interaction. Beyond magnonic applications, our experimental results and theoretical models permit to envisage new functionalities even for the more mature SAW technology.
Here, we put forward the idea of a tunable SAW filter where the SAW phase and/or the resonance frequency can be controlled by a judicious and fine orientation of a permanent magnet. 
For instance, microelectromechanical systems could  be exploited to finely control the orientation of the magnetic field with respect to the propagation direction of the SAW, in order to modify the SWs properties and  control the dynamic magnetoelastic coupling. 
Our velocity changes are still moderate (few 10$^{-5}$) but comparable with phase tunable SAW device on LiNbO$_3$ substrate \cite{Kao2004}. However by increasing the film thickness and by adopting materials with enhanced magnetoelastic and piezoelectric couplings (e.g. TbCo$_2$/FeCo thin film on LiNbO$_3$ in Ref.~\onlinecite{Zhou2014})  the frequency shift can be increased by several orders of magnitude. 

\section{Appendix A}

The purely magnetic free energy density, $f$, normalized to the saturation magnetization ($M_s$), can be expressed as the sum of four terms
\begin{equation}\label{magnetic}
f=f_Z+f_1+f_u+f_d
\end{equation}
where the Zeeman ($f_Z$), cubic ($f_1$), uniaxial out-of-plane ($f_u$), and dipolar ($f_d$) contributions read, respectively: 
\begin{eqnarray}\label{f_local}
f_{Z}&=&-\mu_0 H  \Big[
\sin{\theta}\sin{\theta_H}\cos{(\varphi-\psi)}
+\cos{\theta}\cos{\theta_H} 
\Big] \cr\cr
f_{1}&=&\frac{1}{2}\mu_0  H_{1}\Big[ \sin^2{\theta}\cos^2{\theta}
+\sin^4{\theta}\frac{1+ \cos({4\varphi})}{8}\Big]
\cr\cr
f_{u}&=& \frac{1}{2}\mu_0 H_u  \sin^2{\theta}
\cr\cr
f_{d}&=& -\frac{1}{2}\mu_0 \bf{h}_d \cdot \bf{m}
\end{eqnarray}

where $\mu_0H_1={2K_1}/{M_s}$ and $\mu_0H_u={2K_u}/{M_s}$ are effective magnetic fields associated with the cubic anisotropy constant ($K_1$) and the uniaxial out-of-plane anisotropy constant ($K_u$), respectively. The dipolar field, $\bf{h}_d$, in general does not take a closed form \cite{KS1986}
\footnote{For zero wavevector the dipolar free energy takes the simple expression \cite{Hernandez-Minguez2020,Stamps_Hillebrands}: $f_d=-\frac{1}{2}\mu_0 M_s \sin^2 \theta$.}.

The equilibrium direction of the magnetization is obtained by imposing the vanishing of the first derivatives of the purely magnetic free energy density, $f$, with respect to polar and azimuthal angle $\theta$ and  $\varphi$,  respectively ($f_{\theta}=0$ and $f_{\varphi}=0$). 
When the external magnetic field is applied in plane ($\theta_H=\pi/2$), we assume the easy-plane dipolar anisotropy energy of the film to be stronger than the out-of-plane uniaxial anisotropy, so that the first condition  ($f_{\theta}=0$) is fulfilled by $\theta_0=\pi/2$, meaning that also the equilibrium magnetization lies in the film plane.  The second condition ($f_{\varphi}=0$) provides $\varphi_0$, i.e. the equilibrium azimuthal angle of the in-plane magnetization with respect to the [110] axis, as the solution of the equation
\begin{equation}\label{equilibrium}
H\sin{(\varphi_0-\psi)}-\frac{1}{4} H_{1}\sin{(4\varphi_0)}=0 
\end{equation}
In the absence of the external field ($H=0$), one has that for $K_1>0$ the equilibrium magnetization is parallel to [100]  (i.e., $\varphi_0=\pi/4$, see Fig.~\ref{fig:dispo}). 
If the external field is applied parallel to the hard axis, $\mathbf{H}$ // [110] (i.e., $\psi=0$), the equilibrium angle $\varphi_0$ gradually decreases from $\pi/4$ to 0, when $H$ is increased from 0 to $H_1$. Finally, for $H>H_1$ the equilibrium magnetization remains parallel to the external field (i.e., $\varphi_0=\psi=0$).

\section{Appendix B}
\label{appB}

Brillouin light scattering (BLS) and broadband ferromagnetic resonance (BB-FMR) measurements were performed in order to quantitatively evaluate the magnetic parameters of the Fe film.

BLS measurements were carried out in the backscattering geometry, focusing about 200 mW of monochromatic light (wavelength $\lambda$ = 532 nm) onto the sample surface. 
The scattered  light was frequency analyzed by a Sandercock-type 3+3-pass tandem Fabry-Perot interferometer. 
BB-FMR measurements were performed using a Vector Network Analyzer (VNA) which measure the S$_{ij}$-parameters as function of the frequencies at a fixed external magnetic field from 850 to 50~mT. The thin film sample is positioned at the center of a coplanar wave guide with a 300~$\mu$m wide center strip. Each S$_{12}$-parameter was normalized by the S$_{12}$-parameter at a reference field of 150~mT.

The experimental data were analyzed by using a theory of dipole-exchange spin waves, propagating in plane in a tangentially magnetized thin ferromagnetic ﬁlm, presented in detail in previous works \cite{KS1986,Tacchi2019}.  

The best agreement with the experimental data was obtained using the following magnetic parameters: gyromagnetic factor \cite{Stollo2007} $\frac{\gamma}{2\pi}=29.4$ GHz/T, saturation magnetization $M_s=1.7\times 10^6$ A/m, exchange coupling $A_{ex}=2.0\times 10^{-11}$ J/m, cubic anisotropy $K_1=4.96\times 10^4$ J/m$^3$, and out-of-plane uniaxial anisotropy $K_u=1.0\times 10^5$ J/m$^3$. In addition the fit of BLS experimental data provided an accurate estimate not only of the magnetic parameters, but also of the film thickness: $d=54$ nm (a slightly smaller value than in Ref.~\onlinecite{Duquesne2019a}). 

\begin{figure}
    \centering
    \includegraphics[width=\linewidth]{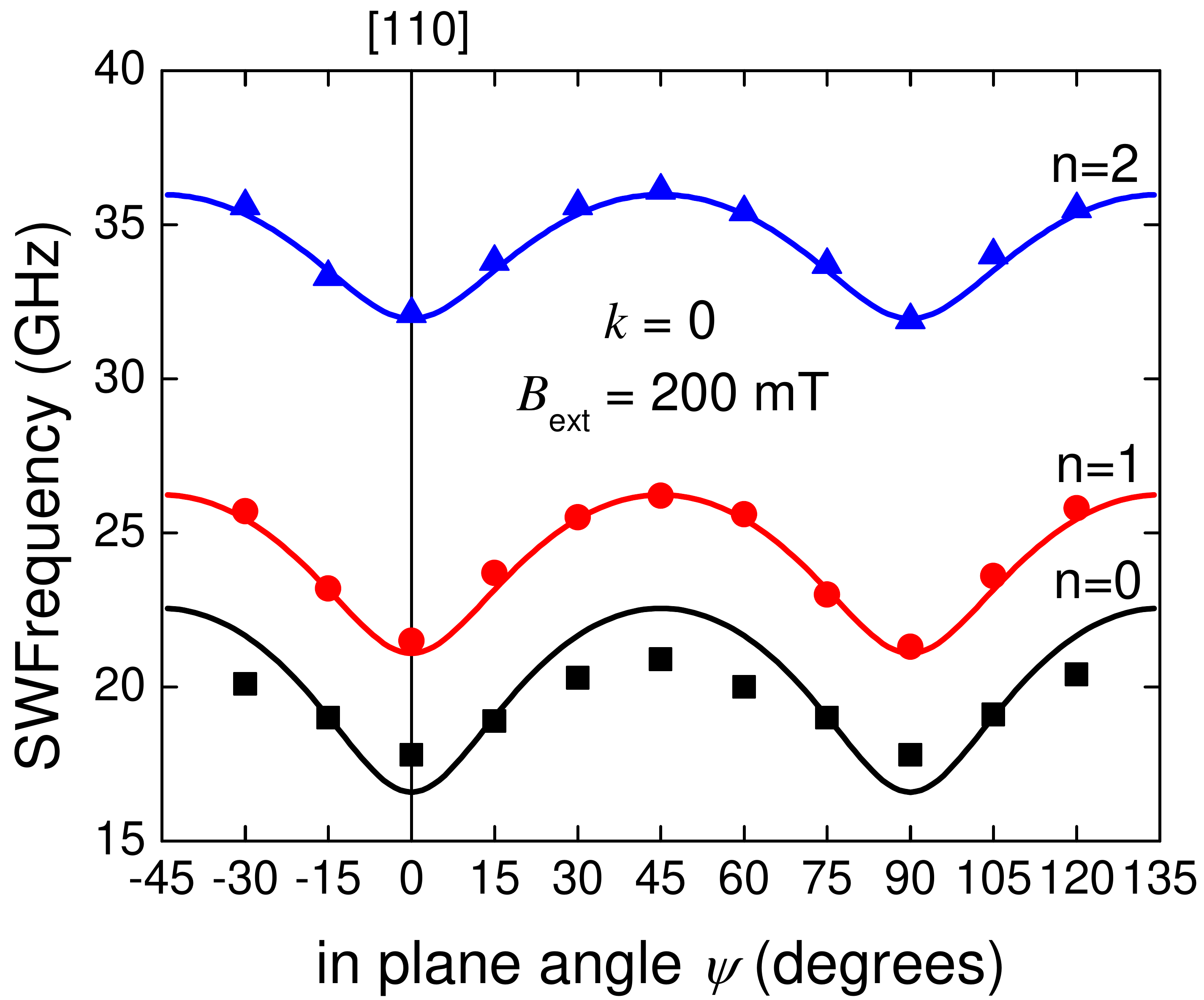}
    \caption{Angular dependence of the three lowest-frequency SW modes ($n=0,1,2$), measured (symbols) by BLS and calculated (lines) using  dipole-exchange SW theory \cite{Tacchi2019} for a Fe film with thickness $d=54$ nm, at zero in-plane wavevector ($k=0$)  and fixed intensity (${B}_{\rm ext}=200$ mT) of the  in-plane applied magnetic field. $\psi$ is the angle formed by $\mathbf{B}_{\rm ext}$ with respect to [110].   }
    \label{fig:B0}
\end{figure}

Figure \ref{fig:B0} shows the dependence of the SW frequency as a function of the in-plane angle $\psi$ between the in-plane applied magnetic field and the [110] Fe direction, measured by BLS at $\textbf{k}$= 0  rad/m for ${B}_{\rm ext}=200$ mT. 
As it can be seen, the sample exhibits the expected bulk Fe cubic anisotropy, whose easy and hard axis are along the [100] and [110] directions, respectively. The lowest frequency mode corresponds to the uniform mode ($n=0$), while the second and the third mode can be identified as the first and second perpendicular standing spin wave mode characterized by $n=1$ and $n=2$ nodes across the film thickness.

\begin{figure}
    \centering
    \includegraphics[width=\linewidth]{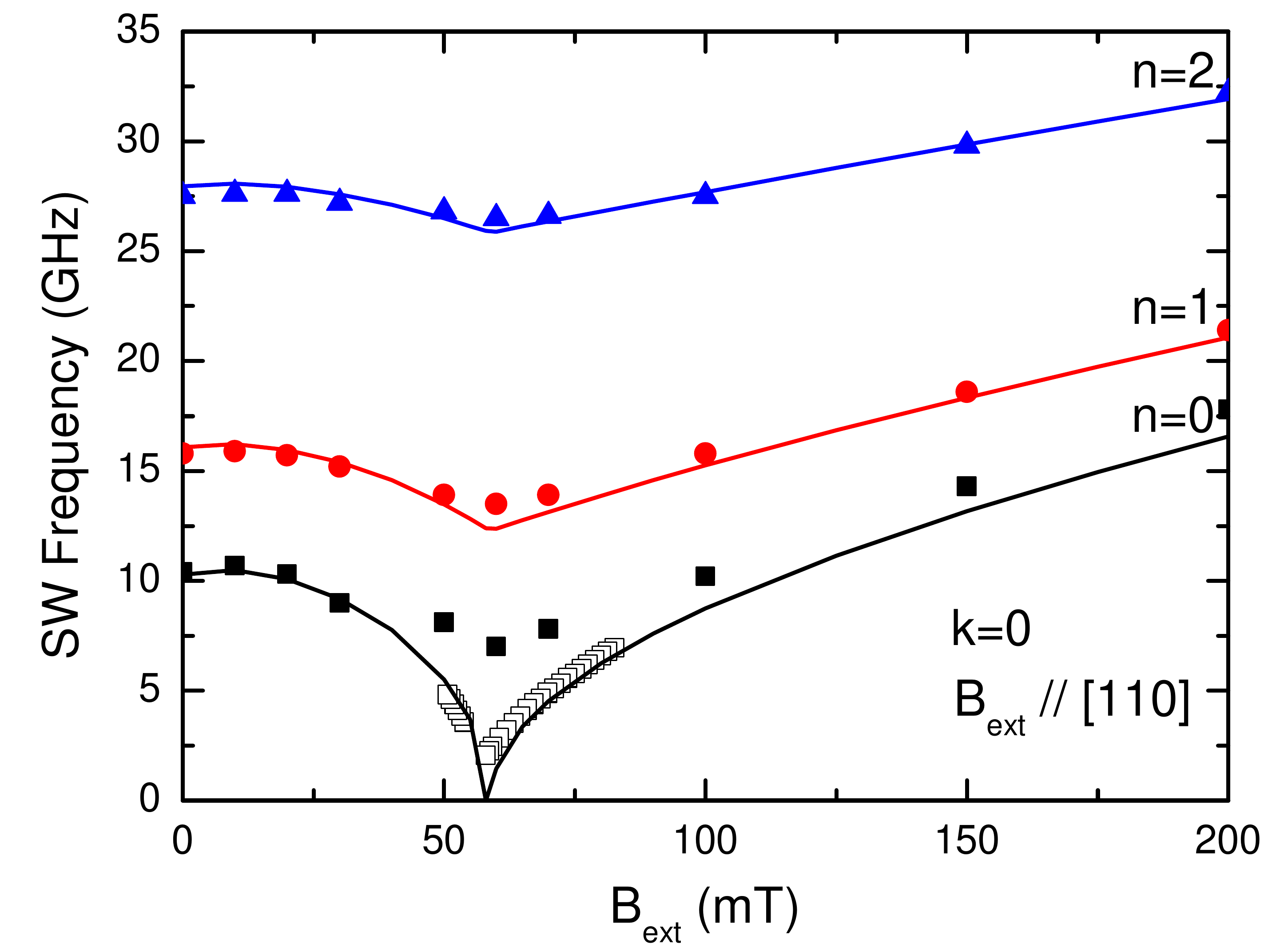}
    \caption{Dependence of the three lowest-frequency SW modes ($n=0,1,2$) as a function of the external field magnitude, measured by BLS (full symbols) and BB-FMR (open symbol) at zero in-plane wavevector ($k=0$) and calculated (lines) using dipole-exchange SW theory \cite{Tacchi2019} for a Fe film with thickness $d=54$ nm. The magnetic field is applied in plane along the hard [110] axis.}
    \label{fig:B1}
\end{figure}

In Figure \ref{fig:B1}, we report the SW frequencies measured by means of BLS and BB-FMR as a function of the magnitude of an external magnetic field, applied along the in-plane [110] hard axis. Note that the mode observed in BB-FMR measurements is in quite good agreement with the lowest frequency mode measured by BLS. All the modes are characterized by a non-monotonic behavior with a local minimum at about 58 mT. This is a typical hard-axis behavior, and indicates a reorientation of the magnetization towards the nearest easy axis when the intensity of the applied field is reduced from saturation. Note that the frequency of the lowest frequency mode is observed to reach an almost zero value in the BB-FMR measurements around ${B}_{\rm r}=58$ mT, while it retains a frequency of about 7 GHz in the BLS ones. This difference can be attributed to a small misalignment of the applied field with respect to the hard axis  \cite{Duquesne2019a}  in the BLS measurements. 

\begin{figure}
    \centering
    \includegraphics[width=\linewidth]{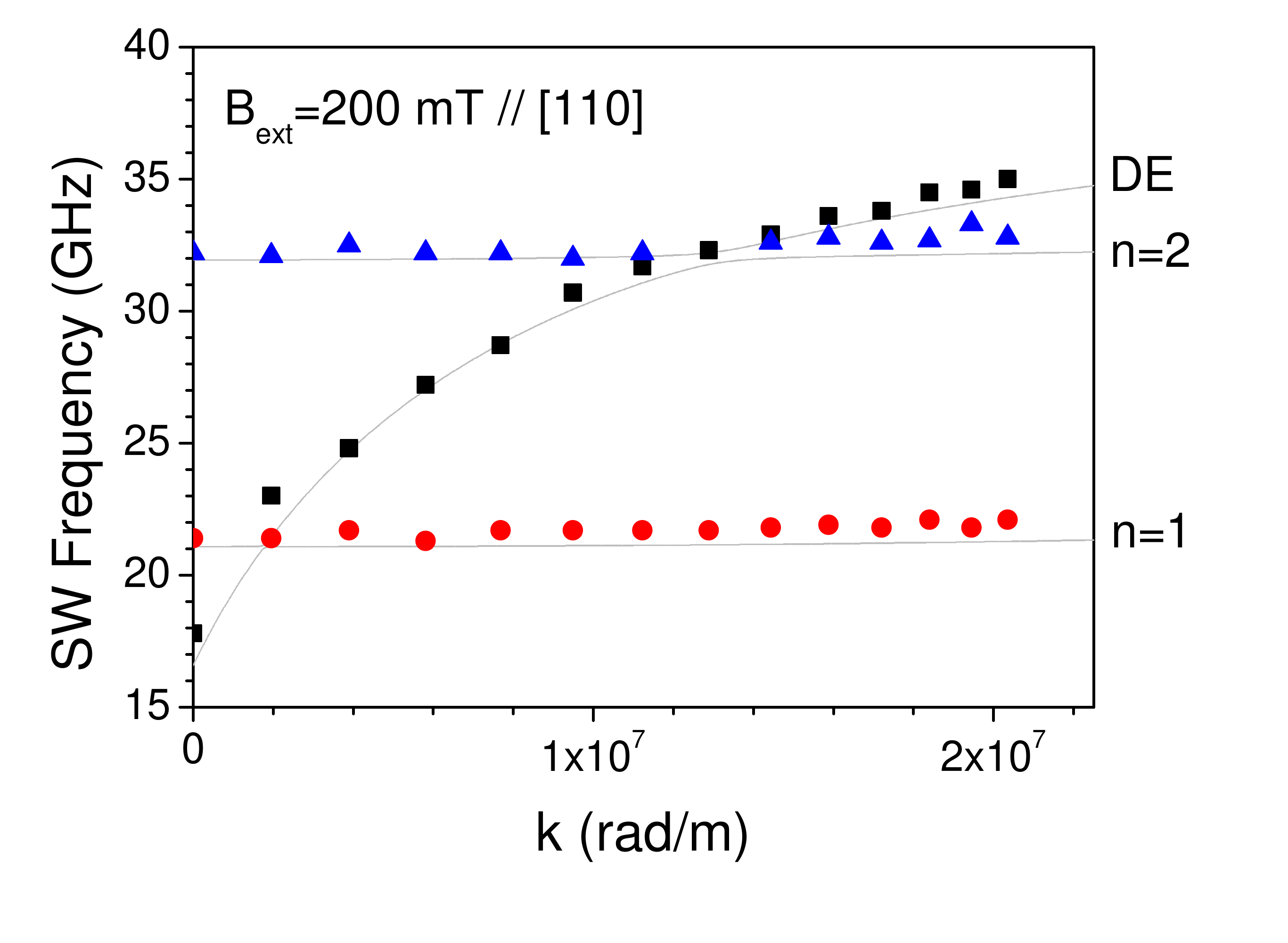}
    \caption{SW dispersion  measured (symbols) by BLS and calculated (lines) using  dipole-exchange SW theory \cite{Tacchi2019} for a Fe film with thickness $d=54$ nm. A magnetic field of fixed intensity (${B}_{\rm ext}=200$ mT) is applied in plane along the hard [110] axis. }
    \label{fig:B2}
\end{figure}

The SW dispersion measured by BLS is shown in Fig.~\ref{fig:B2}. Measurements were performed in the Damon-Eshbach (DE) conﬁguration applying a magnetic field of ${B}_{\rm ext}=200$ mT along the [110] hard axis, and sweeping the in-plane wave vector along the perpendicular direction. Due to the conservation of the in-plane momentum in the scattering process, the wave-vector magnitude k is linked to the incidence angle of light  $\theta$ by the relation $\text{k} = ({4}\pi/\lambda)\sin \theta$. Note that the product of film thickness ($d=54$ nm) and wavevector range  (k $< 2\times 10^7$ rad/m) is not negligible: therefore, the spin-wave frequencies in Fig.~\ref{fig:B2} were calculated taking into account the hybridization between the modes \cite{Tacchi2019}. As it can be seen, the first ($n=1$) and second  ($n=2$) perpendicular standing spin wave modes are characterized by a dispersionless behavior. On the contrary, the lowest frequency mode ($n=0$) exhibits a positive dispersion typical of the DE mode. In fact, in the DE geometry for wave vectors different from zero, the uniform mode becomes the DE mode, which is mainly localized at the film surfaces. 

\section{Appendix C} 
\begin{figure}[h!]
    \centering
    \includegraphics[width=\linewidth]{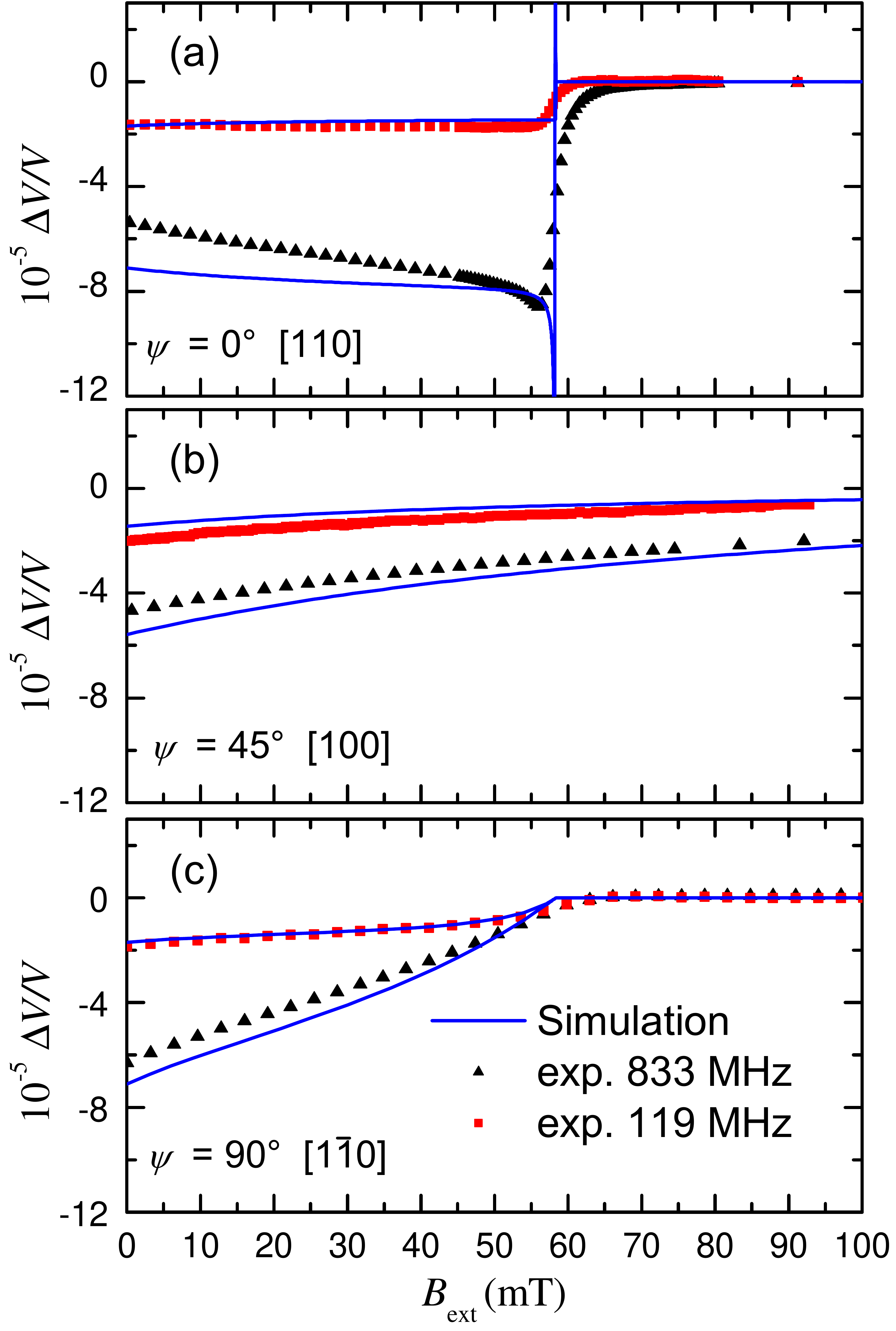}
    \caption{The same as in Fig.\ref{fig:exp2}. Here, calculations consider purely longitudinal waves and the correct SWs dispersion with a Gilbert constant $\alpha = 0.005$.} 
    \label{fig:C}
\end{figure}

As already pointed out by Dreher \textit{et al.} \cite{Dreher2012}, a much simpler description 
of the magnetic field dependence of SAW propagation can be obtained by considering longitudinal waves and by neglecting both strain and rotation terms, i.e. $h_\theta$ =~0 in Eq.~\ref{muhtheta}. This is shown in Fig. \ref{fig:C} where a good agreement between experimental results and this simplified approach is found for $R_{\rm eff} = \nu_{\rm SAW}/24000$ ($\nu_{SAW}$ in MHz). A comparison with Fig.~\ref{fig:exp2} attests that the longitudinal waves approach is sufficient to describe the observed velocity change.

\section{Appendix D}

In Tab.~\ref{tab:elastic}  the elastic constants of Fe and GaAs are reported in the standard and rotated frame.\\

\begin{table}[ht]
    \centering
    \begin{tabular}{|c|c|c|c|c|c|c|c|}
        \hline
        & & \multicolumn{3}{c|}{Standard} & \multicolumn{3}{c|}{Rotated}\\ 
        \hline
         & $\rho$ & C$_{11}$ & C$_{12}$ & C$_{44}$ & C'$_{11}$ & C'$_{12}$ & C'$_{44}$\\
         \hline
        Fe & 7851 & 230.4 & 134.1 & 115.9 & 298.15 & 66.35 & 115.9  \\
        \hline
        GaAs & 5317 & 118.4 & 53.7 & 59.1 & 145.15 & 26.95 & 59.1 \\
        \hline
    \end{tabular}
    \caption{Mass density (in kg.m$^{-3}$) and elastic constants (in GPa) and  of Fe \cite{Adams2006} and GaAs \cite{Cottam1973}.}
    \label{tab:elastic}
\end{table}

Tab.~\ref{tab:epsilon} displays characteristics of the surface acoustic waves. Some values are needed to carry on calculations based on Eq.~\ref{eqP},~\ref{eq3} and~\ref{DeltaP}.\\

\begin{widetext}
\begin{table}[ht]
    \centering
    \begin{tabular}{|l|c|c|c|c|}
        \hline
        Frequency (MHz) 					&  $119$			& $357$ 		& $595$ 	& $833$			\\
        \hline
        $v_{\mathrm{SAW}}$~~ (m.s$^{-1}$) 				& $2856.24$			& $2862.12$		& $2866.20$	& $2868.78$		\\
        \hline
        $<|\varepsilon_{XZ}|^{2}>$  				& \val{1.60}{-11}	& \val{4.11}{-10}	& \val{1.77}{-9}	&  \val{4.54}{-9}		\\	
        \hline
        $<|\varepsilon_{XX}|^{2}>$  				& \val{5.61}{-7}		& \val{1.52}{-6}		& \val{2.31}{-6}		&  \val{2.97}{-6}	\\	
        \hline
        $<\varepsilon_{XX}~.~\overline{\varepsilon_{XZ}}>$	& \val{-2.57}{-9}$\times i$	& \val{-2.14}{-8}$\times i$	& \val{-5.45}{-8}$\times i$	&  \val{-9.82}{-8}$\times i$	\\	
        \hline
        $<|\omega_{XZ}|^{2}>$					& \val{1.06}{-6}		& \val{3.20}{-6}		& \val{5.37}{-6}		& \val{7.58}{-6}		\\	
        \hline
        $<\varepsilon_{XZ}~.~\overline{\omega_{XZ}}>$		& \val{-3.55}{-9}	& \val{-3.14}{-8}	& \val{-8.44}{-8}	& \val{-1.61}{-7}	\\	
        \hline
        $<\varepsilon_{XX}~.~\overline{\omega_{XZ}}>$		& \val{7.71}{-7}$\times i$	& \val{2.21}{-6}$\times i$	& \val{3.52}{-6}$\times i$	& \val{4.73}{-6}$\times i$	\\	
        \hline
        $<|u_{X}|^{2}>$	~~ (m$^{2}$)				& \val{8.18}{-18}	& \val{2.48}{-18}	& \val{1.36}{-18}	&  \val{8.90}{-19}	\\
        \hline
        $<|u_{Z}|^{2}>$	~~ (m$^{2}$)				& \val{1.56}{-17}	& \val{5.30}{-18}	& \val{3.25}{-18}	&  \val{2.37}{-18}	\\
        \hline
        $<|u_{X}|>$	~~~ (m)					& \val{2.86}{-9}	& \val{1.57}{-9}	& \val{1.16}{-9}	&  \val{9.42}{-10}	\\
        \hline
        $<|u_{Z}|>$	~~~ (m)					& \val{3.94}{-9}	& \val{2.30}{-9}	& \val{1.80}{-9}	&  \val{1.54}{-9}		\\
        \hline
        $<|\varepsilon_{XX}|>$					& \val{7.48}{-4}	& \val{1.23}{-3}	& \val{1.52}{-3}	&  \val{1.72}{-3}		\\
        \hline
        $<|\varepsilon_{XZ}|>$					& \val{3.45}{-6}	& \val{1.75}{-5}	& \val{3.65}{-5}	&  \val{5.85}{-5}		\\
        \hline
        $<|\omega_{XZ}|>$					& \val{1.03}{-3}  	& \val{1.79}{-3}	& \val{2.32}{-3}	& \val{2.75}{-3}		\\
        \hline
        $P_{ac}^{layer}$ ~~~~(W)					& \val{8.04}{-3}	& \val{2.38}{-2}	& \val{3.91}{-2}	&  \val{5.44}{-2}		\\
        \hline
         $P_{ac}^{total}$ ~~~~~(W)					& 1			& 1			& 1 			& 1				\\
        \hline
    \end{tabular}
    \caption
    {
    Elastic parameters used in calculations based on Eq.~\ref{eqP}, \ref{eq3} and \ref{DeltaP}.
    Angle brackets stand for the mean value over the thickness of the iron layer. The values are normalized by imposing a total acoustic power of 1~W.
    Calculations are not affected by this arbitrary normalization.
    $P_{ac}^{total}$ is the total acoustic power carried by the surface wave.
    $P_{ac}^{layer}$ is the acoustic power carried by the surface wave only in the iron layer.
    }
    \label{tab:epsilon}
\end{table}
\end{widetext}

In Tab.~\ref{tab:magnetic}  the magnetic parameters of Fe used in previous calculations are reported.\\

\begin{table}[ht]
    \centering
    \begin{tabular}{|l|c|}
        \hline
        Gyromagnetic factor &  $\gamma$ = 1.847 10$^{11}$ Hz.T$^{-1}$\\ 
        \hline
        Saturation magnetization & M$_S$ = 1.7 10$^6$ A.m$^{-1}$\\
        \hline
        Exchange coupling & A$_{ex}$ = 2.0 10$^{-11}$ J.m$^{-1}$\\
        \hline
        Cubic anisotropy & K$_{1}$ = 4.96 10$^{4}$ J.m$^{-3}$\\
        \hline
        Out of plane uniaxial anisotropy & K$_{u}$ = 1.0 10$^{5}$ J.m$^{-3}$\\
        \hline
        Magnetoelastic constant & B$_{2}$ = -7 10$^{6}$ J.m$^{-3}$\\
        \hline
    \end{tabular}
    \caption{Magnetic parameters of Fe.}
    \label{tab:magnetic}
\end{table}

\begin{acknowledgments}
The authors acknowledge C. Gourdon and L. Thevenard for fruitful discussions and careful reading of the manuscript.
They acknowledge the staff of the MPBT (physical properties low temperature) platform of Sorbonne University for their support as well as L. Becerra and M. Rosticher for optical and electronic lithography and A. Anane for dry etching.
The authors acknowledge support from the Agence National de la Recherche française Grant No. ANR-22-CE24-0015 SACOUMAD and from the European Union within the HORIZON-CL4-2021-DIGITAL-EMERGING-01 Grant No. 101070536 MandMEMS.\\ 
\end{acknowledgments}


%

\end{document}